\documentclass[twocolumn,prl,superscriptaddress,showpacs,preprintnumbers,amsmath,amssymb]{revtex4-1}
\usepackage{graphicx}
\usepackage{color}

\begin{document}

\title{Mermin-Wagner fluctuations in 2D amorphous solids}

\author{Bernd Illing}
\affiliation{Department of Physics, University of Konstanz, D-78457 Konstanz, Germany}
\author{Sebastian Fritschi}
\affiliation{Department of Physics, University of Konstanz, D-78457 Konstanz, Germany}
\author{Herbert Kaiser}
\affiliation{Department of Physics, University of Konstanz, D-78457 Konstanz, Germany}
\author{Christian Klix}
\affiliation{Department of Physics, University of Konstanz, D-78457 Konstanz, Germany}
\author{Georg Maret}
\affiliation{Department of Physics, University of Konstanz, D-78457 Konstanz, Germany}
\author{Peter Keim}
\email{peter.keim@uni-konstanz.de}
\affiliation{Department of Physics, University of Konstanz, D-78457 Konstanz, Germany}

\date{\today}

\begin{abstract}
{
In a recent comment, M. Kosterlitz described how the discrepancy about the lack of broken translational symmetry in two dimensions - doubting the existence of 2D crystals - and the first computer simulations foretelling 2D crystals at least in tiny systems, motivated him and D. Thouless to investigate melting and suprafluidity in two dimensions [Jour. of Phys. Cond. Matt. \textbf{28}, 481001 (2016)]. The lack of broken symmetries proposed by D. Mermin and H. Wagner is caused by long wavelength density fluctuations. Those fluctuations do not only have structural impact but additionally a dynamical one: They cause the Lindemann criterion to fail in 2D and the mean squared displacement not to be limited. Comparing experimental data from 3D and 2D amorphous solids with 2D crystals we disentangle Mermin-Wagner fluctuations from glassy structural relaxations. Furthermore we can demonstrate with computer simulations the logarithmic increase of displacements predicted by Mermin and Wagner: periodicity is not a requirement for Mermin-Wagner fluctuations which conserve the homogeneity of space on long scales.
}
\end{abstract}


\maketitle

For structural phase transitions it is well known that the microscopic mechanisms breaking symmetry are not the same in two and in three dimensions. While 3D systems typically show first order transitions with phase equilibrium and latent heat, 2D crystals melt via two steps with an intermediate hexatic phase. Unlike in 3D, translational and orientational symmetry are not broken at the same temperature in 2D. The scenario is described within the so called KTHNY-theory \cite{Kosterlitz1972,Kosterlitz1973,Halperin1978,Nelson1979,Young1979} which was confirmed e.g. in colloidal mono-layers \cite{Keim2007,Kosterlitz2016}. However, for the glass transition it is usually assumed that dimensionality does not play a role for the characteristics of the transition and 2D and 3D systems are frequently used synonymously \cite{Doliwa2000,Harrowell2006,Shintani2006,Berthier2011,Hunter2012}. In a recent manuscript on the other hand, differences of glassy dynamics in two and in three dimensions are reported. Large scale computer simulations show transient localization to be absent in 2D and translational and orientational correlations to decouple in 2D but not in 3D \cite{Flenner2015}.\\

\begin{figure}[t]
\includegraphics[width=.45\textwidth]{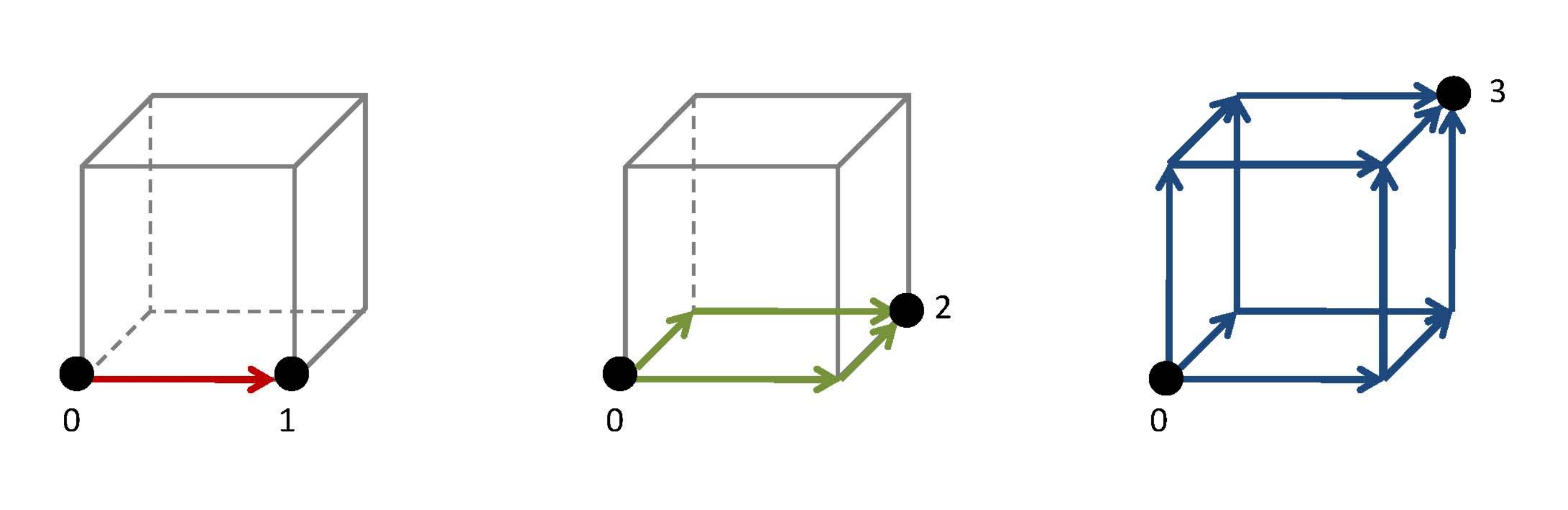}
\caption{\label{Fig01} Counting the path to cover space in various dimensions. In 1D fluctuations can add up independently on an axis along 0 to 1 while in 3D they have to be correlated along the six different ways from 0 to 3 and stay finite. In 2D fluctuations add up logarithmically.}
\end{figure}

In the present manuscript we compare crystal and glass data and show that Mermin-Wagner fluctuations, well known from 2D crystals, are also present in amorphous solids \cite{Mermin1966,Mermin1968}. Mermin-Wagner fluctuations are usually discussed in the framework of long range order (magnetic or structural). But, in the context of 2D crystals they have also impact on dynamic quantities like mean squared displacements. Long before 2D melting scenarios were discussed, there was an intense debate whether crystals and perfect long range order (including magnetic order) can exist in 1D or 2D at all \cite{Bloch1930,Peierls1934,Landau1937a,Landau1937b}. A beautiful heuristic argument given by Peierls \cite{Peierls1934} is as follows: consider a 1D chain of particles with nearest neighbor interaction. The relative distance fluctuation between particle $n$ and particle $n+1$ at finite temperature may be $\xi$. Similar is the fluctuation between particle $n+1$ and $n+2$. Thus, the relative fluctuation between second nearest neighbours, namely particle $n$ and $n+2$ is $\sqrt{2}\cdot\xi$ since they add up statistically independently. Thus the amplitude of the fluctuations grows with $\sqrt{N}\cdot\xi$ if $N$ counts the number of particles in the chain. Periodicity cannot exist at large scales in 1D crystals. To cover 3D space, one has to investigate three linear independent directions. Within a cube for instance there are six ways to get along the space diagonal say, from the lower left front corner to the upper right back corner (see Fig.~\ref{Fig01}). It follows that in 3D the fluctuations cannot add up independently and the amplitude of the fluctuations stays finite being of the order of $\xi$. In 2D one can show that fluctuations add up logarithmically at finite temperatures. Translational correlation functions decay algebraically while, and this is important to note, orientational order is not affected \cite{Mermin1966,Mermin1968,Peierls1934,Frohlich1981,Dash1978}.\\

What is the impact of Mermin-Wagner fluctuations? They are long(est) wavelength density fluctuations and mapping locally a perfect mathematical 2D lattice with commensurable density and orientation, one finds the displacement of particles to diverge. It is shown analytically that this displacement from perfect lattice sites increases in two dimensions logarithmically with distance \cite{Mermin1968,Frohlich1981}. 
Having a closer look at the arguments given in \cite{Peierls1934} one finds that periodicity is not a requirement for those fluctuations. They will also be present in other 2D (and 1D) systems like quasi-crystals or amorphous structures, provided the fact that nearest neighbour distances have low variance (unlike e.g. in a gas).
D. Cassi, F. Merkl, and H. Wagner \cite{Cassi1992,Merkl1994,Cassi1996} mapped the absence of spontaneously broken symmetries to the recurrence probability of random walks. There it is proven, that spontaneous magnetization on amorphous or fractal networks can not occur in $d \leq 2$. The dualism with random walks shows that Mermin-Wagner fluctuations are time dependent for nonzero temperatures. In 2D crystals, Mermin-Wagner fluctuations cause translational correlation functions to decay algebraically \cite{Halperin1978}. With respect to dynamic measures, they cause the mean square displacement (MSD) to diverge and the standard Lindemann parameter to fail. The impact on the dynamics is independent of periodicity and should be found in quasi-crystals and amorphous solids, too \cite{Keim2015}. Using local coordinates as introduced by Bedanov, Gadiyak, Farztdinov, and Lozovik \cite{Bedanov1985,Lozovik1985}, namely subtracting the trajectories of the nearest neighbors, the so called reduced or local MSD stays finite in a 2D crystal but still diverges in the fluid. This defines a dynamic Lindemann criterion in 2D \cite{Zheng1998} which is a maximal threshold for the local displacements in a solid. In the language of glass theory, the nearest neighbors are given by the cage and the ‘cage-relative mean square displacement’ (CR-MSD) was shown to have much more contrast e.g. for dynamical heterogeneities in a 2D glass former compared to standard MSD \cite{Mazoyer2009,Mazoyer2011}. Recent work by Vivek et al. using cage-relative intermediate scattering functions support this idea \cite{Vivek2016} and computer simulations by Shiba et al. independently found similar results \cite{Shiba2016}.\\

\begin{figure}
\centering
\includegraphics[width=.45\textwidth]{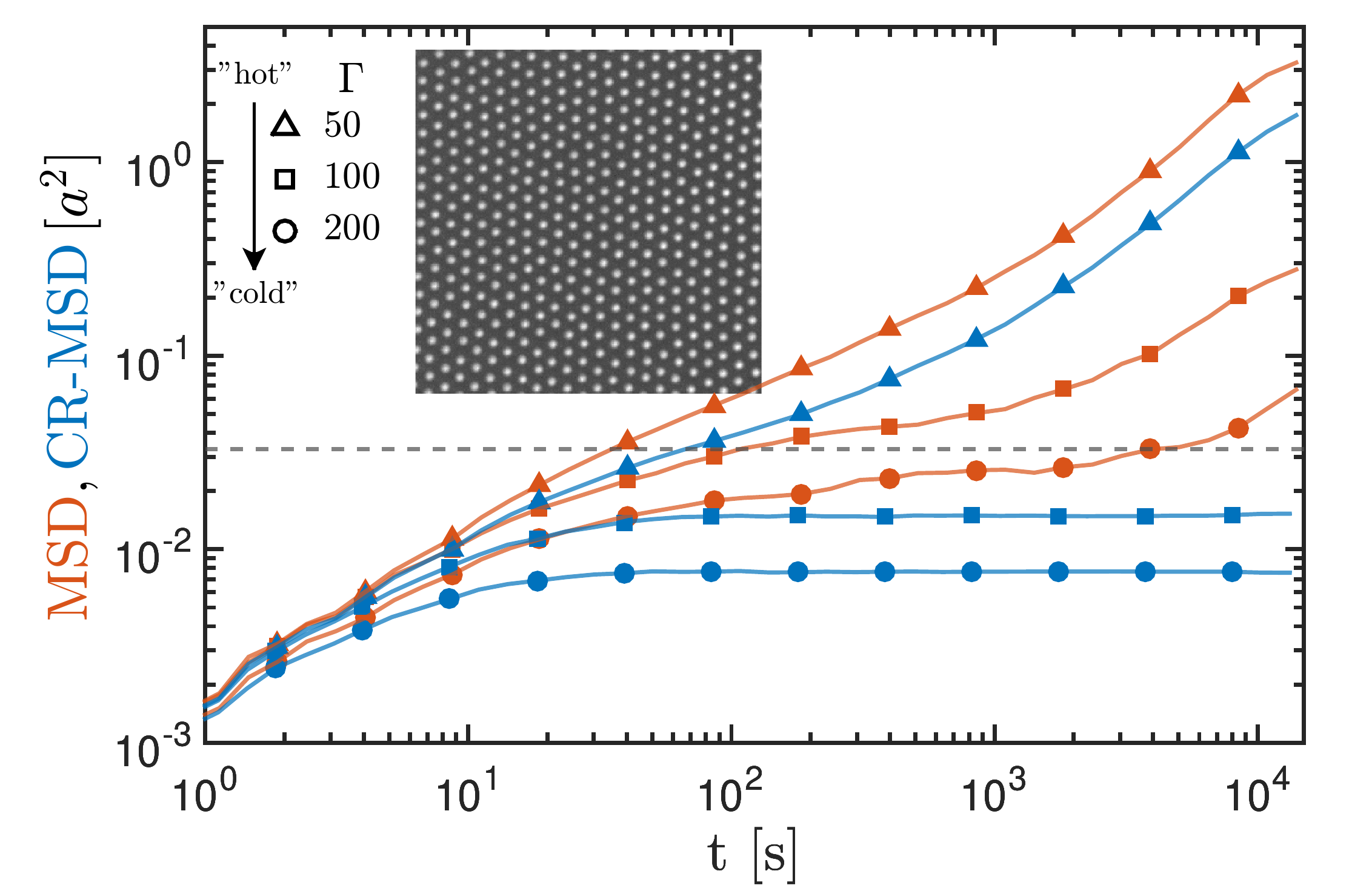}
\includegraphics[width=.45\textwidth]{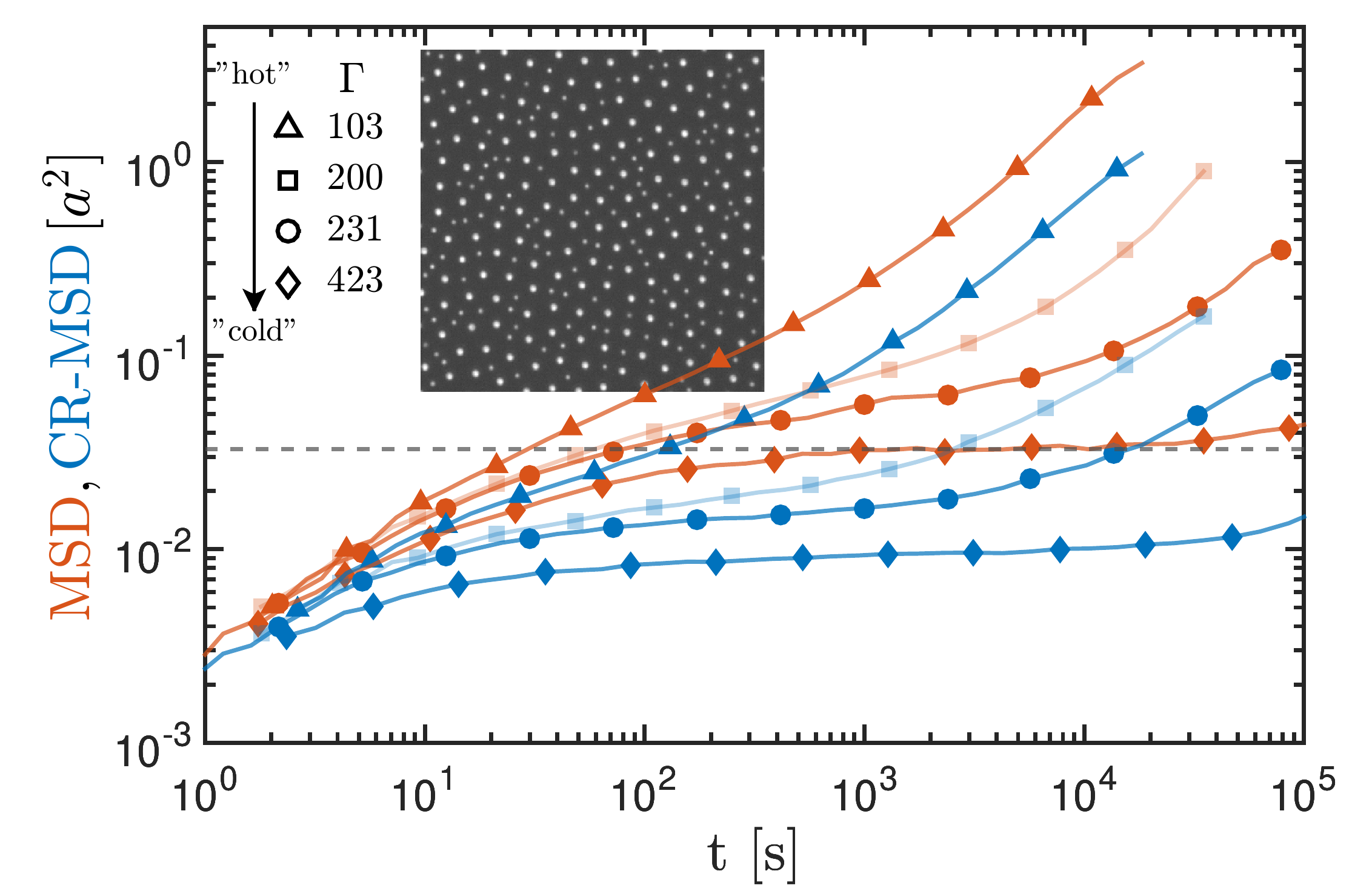}
\includegraphics[width=.45\textwidth]{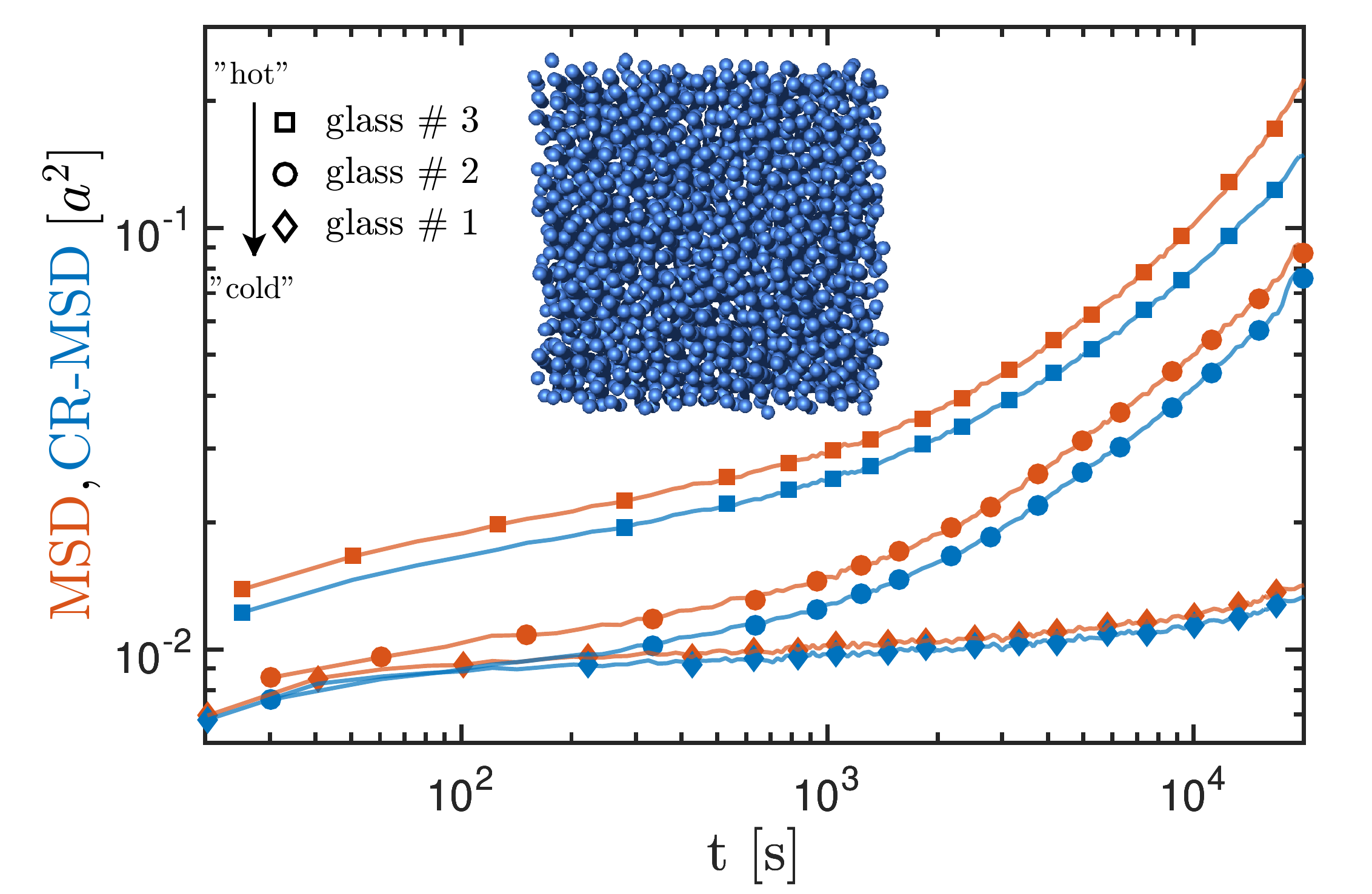}
\caption{\label{Fig02} Upper plot: mean squared displacement of a defect free 2D crystal (red). The control parameter $\Gamma$ is an inverse temperature, increasing $\Gamma$ triggers solidification (melting is at $\Gamma_m=60$). On long times, the MSD diverges. Even a 2D crystal has fluid-like character due to Mermin-Wagner fluctuations, while using local coordinates the cage-relative mean square displacements (CR-MSD) stays finite (blue). Middle: in a 2D glass an additional $\alpha$ -process causes the CR-MSD (blue) not to stay finite and the amplitude of the global fluctuation given by the MSD (red) is significantly larger (the glass transition is at $\Gamma_G \approx 200$, this curve is therefore plotted transparent). Both 2D systems labeled with triangles are fluid. Lower plot: MSD (red) and CR-MSD (blue) for a 3D glass for various supercooling (details below). The difference in the amplitude is significantly smaller and the upturn seem to appear simultaneously. Note that the ordinate is zoomed in in 3D compared to 2D. 2D and 3D glasses labeled with diamonds are deep in the glass phase, and the $\alpha$-process starts to appear at the end of the accessible time window.}
\end{figure}

Fig~\ref{Fig02} shows the mean squared displacement (MSD) where the sum runs over all particles $N$ and the brackets additionally denote an average about starting times $\tau$

\begin{equation}
\langle r^2(t)\rangle = \frac{1}{N} \sum_{j=1}^N [\vec{r}_j(t+\tau)-\vec{r}_j(\tau)]^2
\label{eqn01}
\end{equation}
and cage-relative mean squared displacements (CR-MSD)
\begin{eqnarray}
\label{eqn02}
\langle r^2(t)\rangle^\mathrm{CR} & = & \frac{1}{N} \sum_{j=1}^N  \left[\frac{}{}(\vec{r}_j(t+\tau)-\vec{r}_j(\tau))\right. \nonumber \\
& ~ & - \frac{1}{N_j}\sum_{i=1}^{N_j}\left.\frac{}{}(\vec{r}_i(t+\tau)-\vec{r}_i(\tau))\right]^2
\end{eqnarray}
for crystals and amorphous solids at various temperatures. The second sum in Eqn.~\ref{eqn02} is the center of mass of the cage given by the $N_j$ nearest neighbours of particle $j$ determined by Voronoi-Tesselation. The left plot shows the standard MSD as function of time in red for a fluid system (red triangles) and two crystalline samples (red squares and circles). In a 2D crystal the mean squared displacement is not confined. This indicates the failure of the Lindemann-criterion in 2D. Using cage-relative coordinates (blue curves) the fluid data still diverge (blue triangles) but the CR-MSD from solid samples are confined (blue squares and triangles). The dashed line shows the critical value given by the dynamic Lindemann criterion (which is $\gamma_L = 0,033$ for the given system). Below this value the system is a crystal \cite{Zheng1998,Zahn2000}. Since grain boundaries, which emerge for finite cooling rates during preparation of the sample \cite{Deutschlaender2015b}, might cause some plasticity they are excluded in the analysis \footnote{Since even real 3D mono-crystals (without grain boundaries) incorporate vacancies and interstitials due to entropic reasons, the MSD can strictly spoken not be finite due to defect migration, too.}. This is done by analyzing only particles which have a crystalline environment (six nearest neighbours) for the time of investigation.

The plot in the middle shows the same analysis for a glass forming system. The mean squared displacement of the fluid sample is labeled with red triangles, the transparent square label a sample which is glassy but very close to the transition temperature, while the circles and diamonds represent amorphous solids \cite{Klix2012,Klix2015}. Focussing on the cage-relative mean squared displacements (blue curves) one finds that the amplitude of the local displacements is lower but even for the deepest supercooled amorphous solid (blue diamond) there is an upturn for long times. While for the CR-MSD (blue curves) long wavelength phonons are shortcut and thus invisible, the structural relaxation, which typically appears for glasses, is still visible. The so-called $\alpha$-process which is usually attributed to particles escaping their cage given by nearest neighbors is detectable in glass but not in the crystal. Note that the upturn in MSD (red) appears earlier compared to the CR-MSD (blue) in the 2D glass. The right panel shows a 3D glass which lacks per definition Mermin-Wagner fluctuations. The amplitude of the CR-MSD (blue) is only slightly smaller compared to the standard MSD (red) and the upturn seem to happen simultaneously. Thus only structural relaxation is measured which is shifted beyond the accessible time window for the system deepest in the glass (diamonds). The corresponding insets show typical snapshots of the 2D systems (see experimental details below and in the supplemental information while for the 3D system a sketch is shown, reconstructed from structural data of the amorphous solid.

\section{Colloidal systems}

This difference of global and local fluctuations in 2D is already a hallmark of Mermin-Wagner fluctuations but before we focus in orientational and structural decay, the experimental realization of two-dimensional and three-dimensional systems and details about the simulations are briefly discussed. The 2D systems are well established and we investigated crystallization, defects \cite{Keim2004,Gasser2010,Deutschlaender2014,Lechner2015}, and the glass transition \cite{Mazoyer2011,Klix2012,Klix2015} with this setup. They consist of colloidal mono-layers where individual particles are sedimented by gravity to a flat a water/air-interface in hanging droplet geometry. The colloids are a few microns in size and perform Brownian motion within the plane. The control parameter of the system is the parameter $\Gamma = E_{Pot}/E_{Kin}$ given by the ratio of potential energy of the particles (due to mutual dipolar interaction) and the kinetic energy $E_{Kin} =\propto T$ due to thermal motion. It can be interpreted as an inverse temperature (or dimensionless pressure), thus large values of $\Gamma$ refer to small temperatures and vice versa. The whole monolayer consists of a few hundred thousand of particles and a few thousand are monitored by standard video microscopy and digital image analysis. As shown in the supplemental information, about 2\% of pinned particles on a solid substrate is enough to suppress Mermin-Wagner fluctuations.

The 3D colloidal systems consist of more than a billion of particles, dissolved in an organic solvent with identical mass-density, thus particles do not sediment. The colloids are slightly charged thus the interaction is given by Coulomb-interaction screened by a small amount of counter-ions in the solvent (Yukawa-potential). Monitoring is performed with confocal microscopy, providing 3D images with several thousand particles being tracked in the field of view.
Finite size effects in 2D are additionally investigated with computer simulations, specifically Brownian dynamics simulation of hard discs. To prevent crystallization, a binary mixture of different sizes of discs is used. The phase diagram is controlled by entropy (not temperature) and the control parameter in this systems is solely given by the (area) density of discs in the plane.

Colloidal systems are so called soft matter systems: the interaction energy between particles is of the order (tenth of) $eV$, comparable to atomic or molecular systems. But since length-scales (distances between particles) are about $10^{4}$ to $10^{6}$ times larger, energy densities (and therefore elastic moduli) are smaller by $10^{8}$ to $10^{12}$ in 2D and even $10^{12}$ to $10^{18}$ in 3D. Thus soft matter has a rich variety of excited states at moderate temperatures and thermally induced fluctuations are easily accessible. This offers the unique possibility to measure Mermin-Wagner fluctuations in the laboratory. For atomic systems including Graphene it has been argued that sheets of cosmologic size are necessary to detect any realistic amplitude of Mermin-Wagner fluctuations \cite{Abraham1980,Abraham1981,Thompson2009}.

\section{Structural and orientational decay in 2D and 3D}

\begin{figure*}
\begin{tabular}{cc}
 \includegraphics[width=0.5\linewidth]{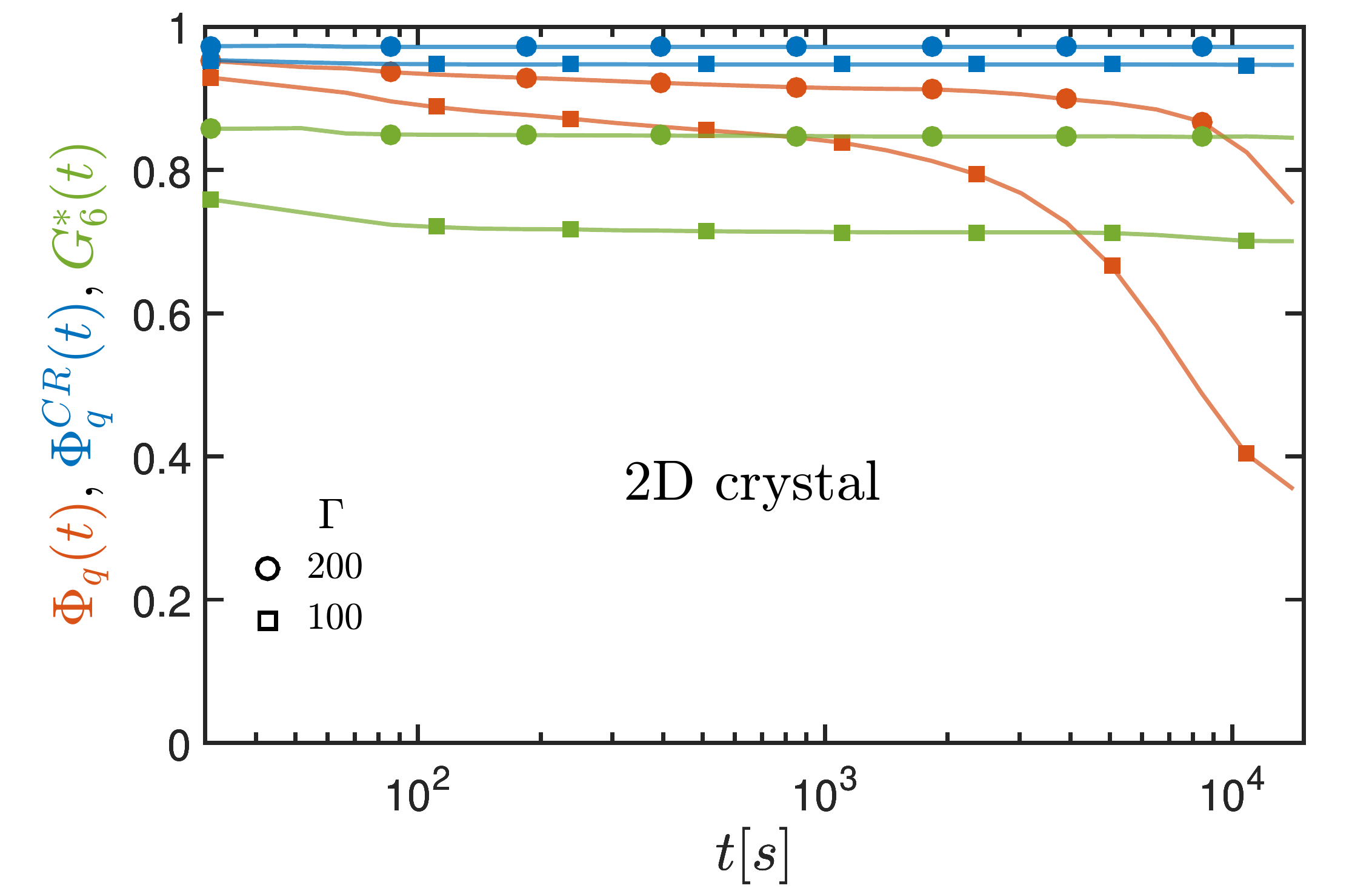} & \includegraphics[width=0.5\linewidth]{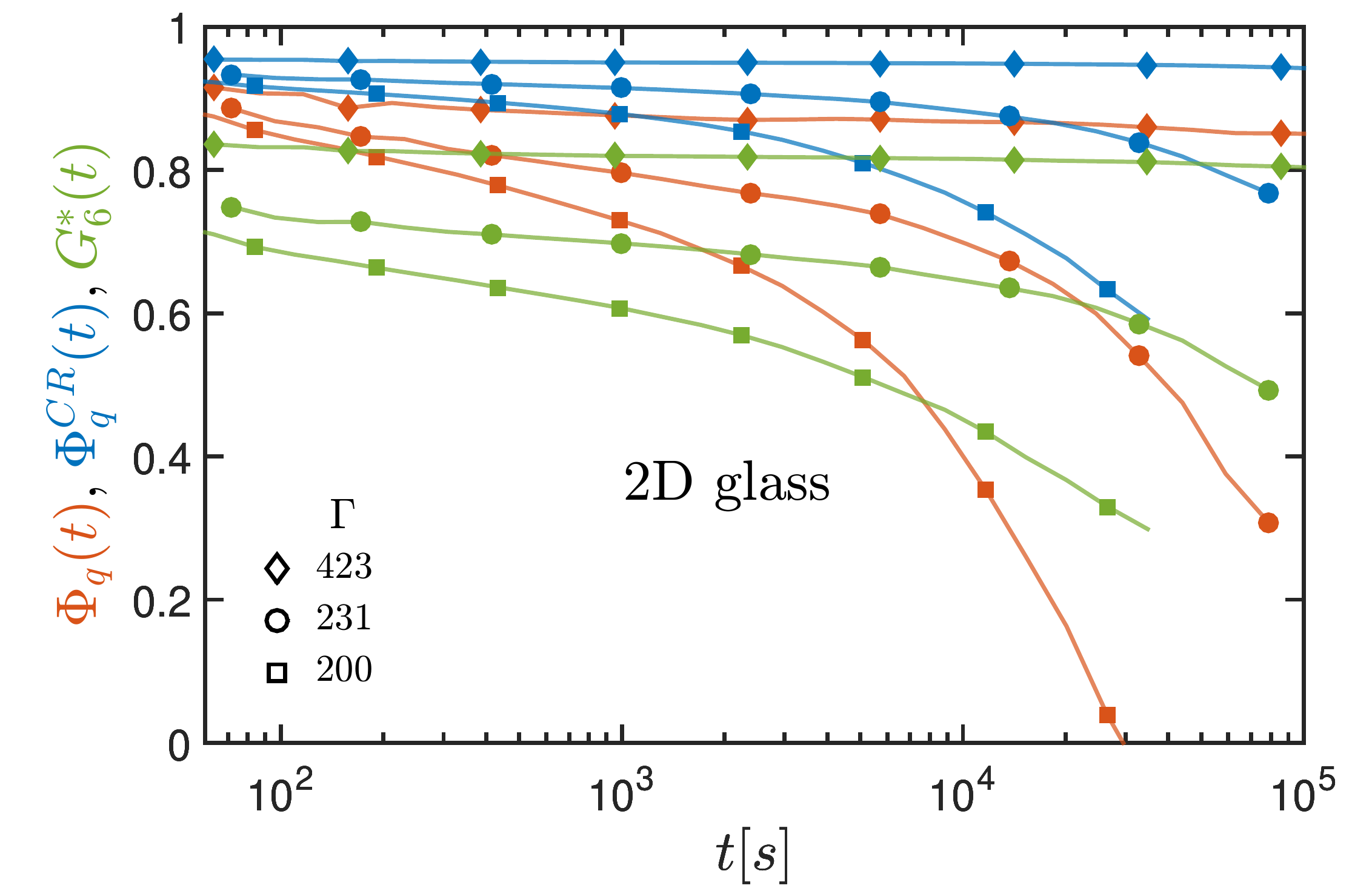}\\
 \includegraphics[width=0.5\linewidth]{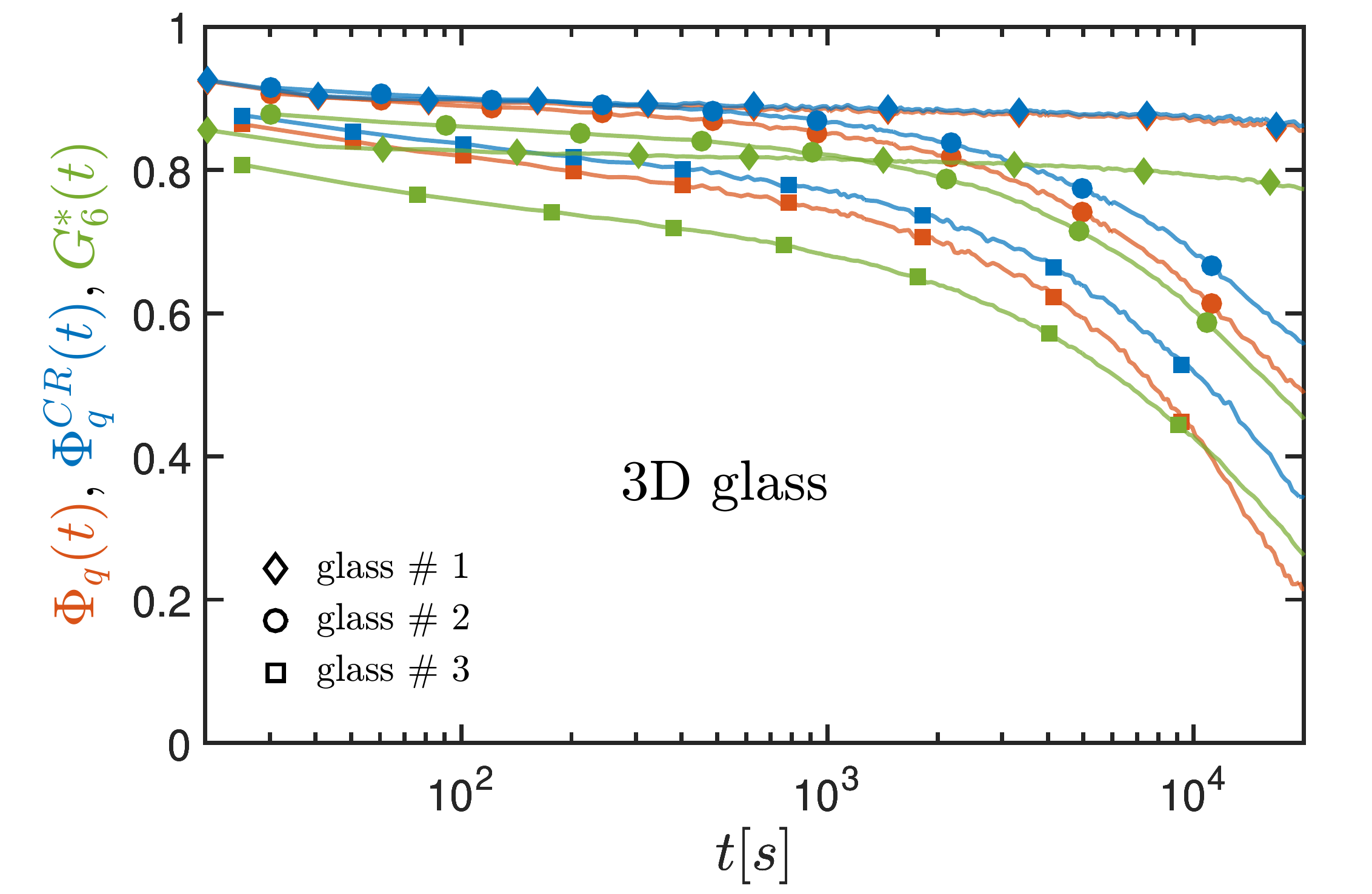} & \includegraphics[width=0.5\linewidth]{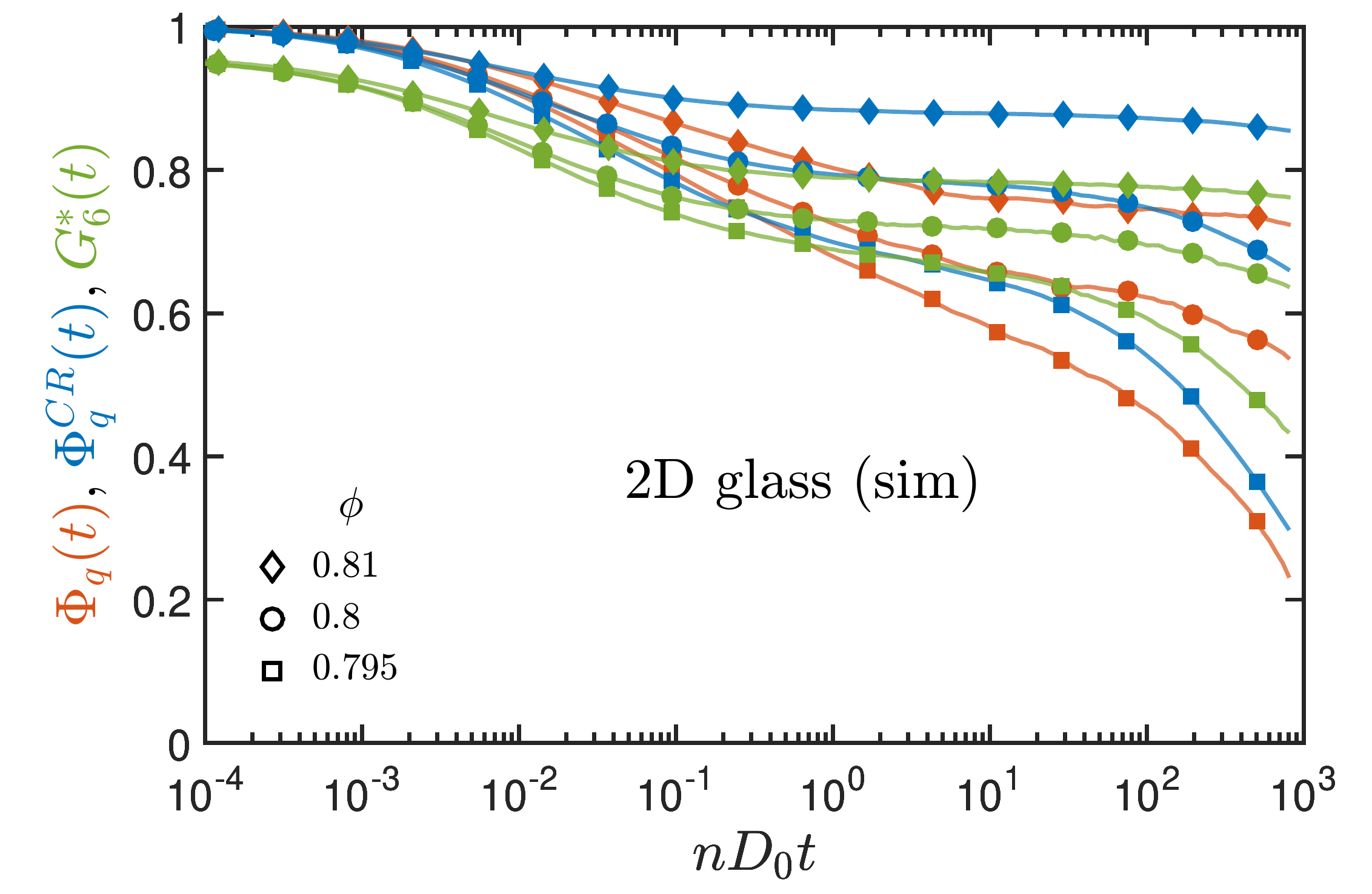} \\
 \end{tabular}
 \caption{\label{Fig03} Self intermediate scattering function $\Phi_q(t)$ (red), Cage-relative self intermediate scattering $\Phi^{CR}_q(t)$ (blue), and bond order correlation function $G_6^*(t)$ (green) for various temperatures of a 2D crystal (upper left), a 2D glass (right column) and 3D glass (lower left). In the 2D crystal $\Phi^{CR}_q(t)$ and $G_6^*(t)$ do not decay, while $\Phi_q(t)$ decays due to Mermin-Wagner fluctuations. In the experimental and simulated 2D glass, $\Phi^{CR}_q(t)$ and $G_6^*(t)$ decay simultaneously due to structural relaxations within the $\alpha$-process, while $\Phi_q(t)$ decays earlier due to Mermin-Wagner fluctuations and the $\alpha$-process. In the 3D glass, only an $\alpha$-process occurs and no separation in timescales is visible. The stiffest glasses (diamonds) do not decay within the accessible time window but indicate the stability of our experiments.}
\end{figure*}

To measure the structural decay and to investigate the $\alpha$-process in glass one frequently uses the intermediate scattering function $\bar{\Phi}(\vec{q},t) = \langle n(\vec{q},t+\tau)\cdot n^\ast(\vec{q},\tau) \rangle$ where $n(\vec{q},t)$ is the fourier-transformation of the density $n(\vec{r},t)$ at time $t$. The self part of the intermediate scattering function $\Phi_q(t)$ ignores cross correlations between different particles at $\vec{r}_k$ and $\vec{r}_l$ but correlates the position of particle $\vec{r}_i$ at $\tau=0$ with its position at $\tau=t$ in Fourier-space,
\begin{equation}
	 \Phi_q(t) = \left \langle \frac{1}{N}\sum_{j=1}^{N} e^{-i\vec{q}\cdot\Delta\vec{u}_j(t)}  \right \rangle  \,  ,
\label{eqn03}
\end{equation}
where the sum runs over all particles $N$. It is nothing but the fourier-transform of the displacements $\Delta u_j(t) = \vec{r}_j(t+\tau)-\vec{r}_j(\tau)$ and the wave-vector $\vec{q}$ is as usual chosen to be $q = 2 \pi /a_0 $. This is where the structure factor $S(q)$ has its first maximum. Angular brackets indicate the canonical ensemble average for simulations and an average about various starting times $\tau$ in experiments. In Figure~\ref{Fig03}, $\Phi_q(t)$ is plotted as red curves for four different systems: the upper ones are the 2D crystal (left) and 2D glass (right) of dipolar particles as in Fig~\ref{Fig02} but omitting the fluid curves. After an initial decay (which is hardly seen on the log-lin scale) due to thermal vibrations the red curves enter a plateau, indicating the dynamic arrest. Only the stiffest glass (diamonds) is stable on the accessible time scale. The lower right plot shows data from simulations of a 2D hard disk system for comparison, where the packing fraction but not temperature is the only control parameter. The qualitative behaviour is the same as for the 2D dipolar glass. The lower left plot shows the 3D glass, again with a typical two step decay, except for the strongest glass (red diamonds) where the decay is hardly visible on the experimental accessible time scale.

In analogy to the CR-MSD one can define a cage relative intermediate scattering function given in blue in Fig.~\ref{Fig03}, where the displacement is reduced by the center of mass motion of the nearest neighbours. In 2D the nearest neighbours are defined by Voronoi-Tessellation while in 3D a cutoff value of $1.2 a_0$ is used to identify particles within the first shell representing the cage \cite{Vivek2016},
\begin{equation}
	 \Phi^\mathrm{CR}_q(t) = \left \langle \frac{1}{N}\sum_{j=1}^{N} e^{-i\vec{q}(\Delta\vec{u}_j(t)-\Delta\vec{u}_j^{cage}(t))} \right \rangle  \, ,
\label{eqn04}
\end{equation}
where the displacement of the cage of particle $j$ given by the $N_j$ neighbours reads $\Delta\vec{u}_j^{cage}(t) = \frac{1}{N_j}\sum_{i=1}^{N_j}(\vec{r}_i(t+\tau)-\vec{r}_i(\tau))$.

We further introduce the bond order correlation function $G_6^*(t) = \langle \psi_{6}^*(t+\tau)\psi_{6}(\tau)\rangle$ which correlates the local director field in time. In the crystal the director field is given by the bond direction to the nearest neighbours in six-folded space:
\begin{equation}
	\psi_{6}=\frac{1}{N_i}\sum_{i}e^{i6\cdot\theta_{ij}(t)} \, ,
\label{eqn05}
\end{equation}
and $\theta_{ij}(t)$ is the time dependent angle of the bond direction between particle $i$ and $j$ and an arbitrary reference axis. For the 2D glass, only $\approx20\%$ of the particles are sixfolded, $\approx75\%$ are five- and seven-folded (together and similar distributed) while $<5\%$ are four- or eight-folded. Thus, for the 2D binary mixture we sum up all relevant director fields,
\begin{equation}
G_6^*(t) = \sum_{n=4}^8 \langle \psi_{n}^*(t+\tau)\psi_{n}(\tau)\rangle \,  ,
\label{eqn06}
\end{equation}
still $G_6^*(t=0)\lesssim 1$. For the 3D glass, the local director field of particle $i$ is given by $Q_{6m}^i = \frac{1}{N_j}\sum_{j=1}^{N_j} q_{6m}(\vartheta_{ij},\varphi_{ij})$ based on the spherical harmonics $q_{lm}(\vartheta,\varphi)$ for $l=6$ with polar $\vartheta$ and azimutal $\varphi$ angle of the bond \cite{Steinhardt1983}. The 3D correlation function reads
\begin{equation}
 G_6^*(t) =  \frac{4\pi}{2l+1}\sum_i^N\sum_{m=-6}^{m=6}Q_{6m}^i(t+\tau)(Q_{6m}^i(\tau))^* \,  ,
\label{eqn07}
\end{equation}
In Figure~\ref{Fig03} we now compare the cage-relative intermediate scattering function $\Phi_q(t)$ plotted in blue and the bond order correlation function $G_6^*(t)$ plotted in green. In the crystal both correlation functions do not decay. As for the MSD, the Mermin-Wagner fluctuations are shortcut using local coordinates. The orientational order does not decay since the modulus of rotational stiffness, (usually called Frank's constant in analogy to liquid crystal theory) is infinite in a 2D crystal \cite{Nelson1979,Keim2007} even if translational order decays and the MSD diverges. Long range bend and splay are suppressed while long range density fluctuations are allowed in 2D crystals \cite{Nelson1979,Eisenmann2005}.

For the soft glasses $\Gamma = 200/231$ (green and blue squares/circles in upper right Fig~\ref{Fig03}) both correlation function $\Phi^\mathrm{CR}_q(t)$ and $G_6^*(t)$ decay but not the stiffest one for $\Gamma=423$ (diamonds) where the standard $\Phi_q(t)$ was already stable within the given time window. Note that the timescales for orientational and cage-relative structural decay is the same for identical $\Gamma$ (comparing curves with green and blue squares for $\Gamma=200$ and green and blue circles for $\Gamma=231$). The separation in timescales compared to the standard structural decay $\Phi_q(t)$ (red) is clearly visible. The 2D simulations show the same behaviour which means 2D glasses are affected by slow Mermin-Wagner fluctuations AND structural relaxations. The lower left plot shows the 3D glass. The stiffest glass\#1 (diamonds) is almost stable. In glass\#2 (blue, red, and green circles) and glass\#3 (blue, red, and green squares) all correlation functions decay on the same timescale due to structural relaxations but without Mermin-Wagner fluctuations. We conclude that 2D crystals are affected by Mermin-Wagner fluctuations, 2D glasses are affected by Mermin-Wagner fluctuations and $\alpha$-relaxation while 3D glasses are only affected by $\alpha$-relaxation.

\section{Finite size effects}

\begin{figure}[h]
\includegraphics[width=.5\textwidth]{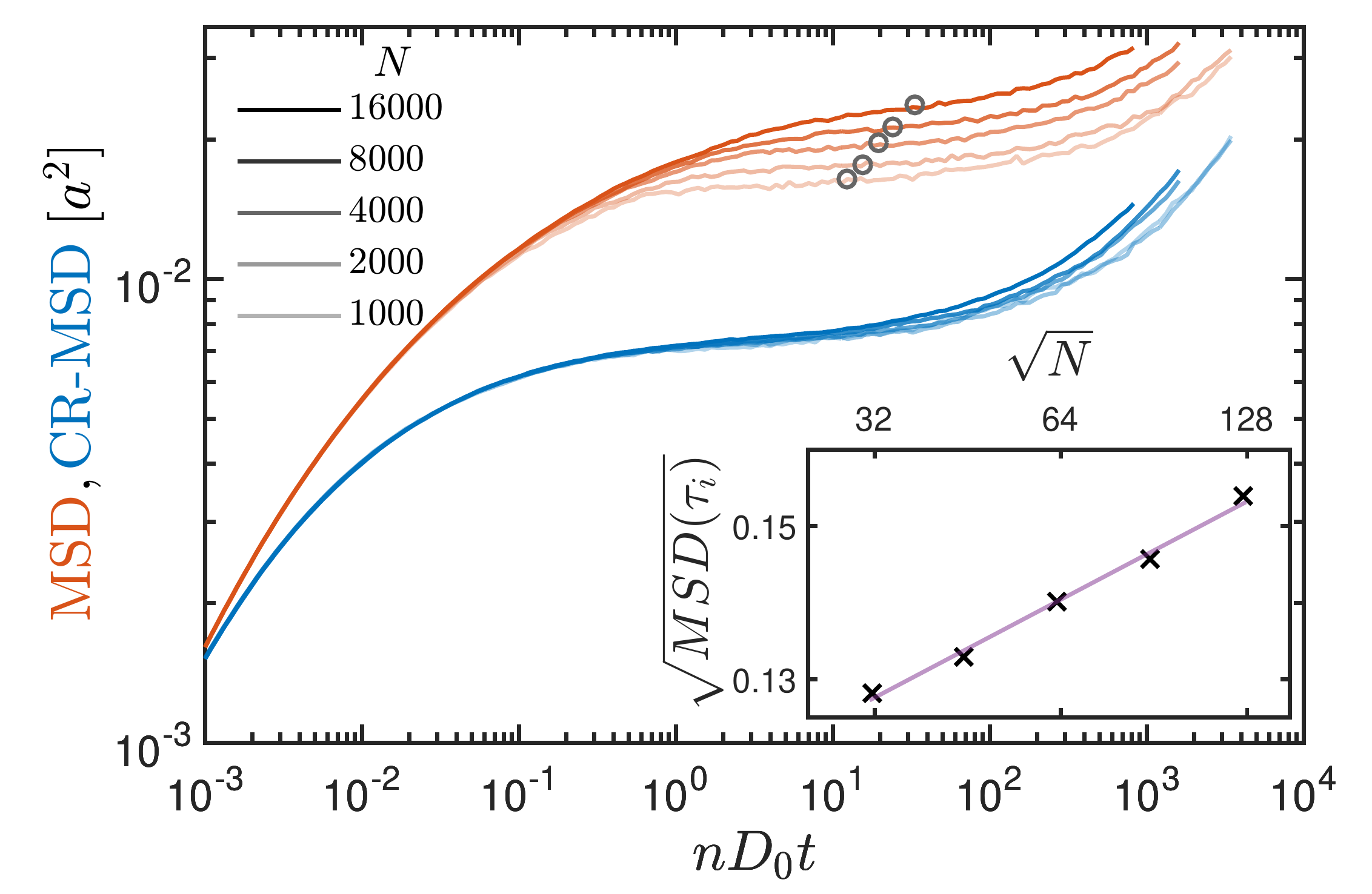}
\caption{\label{Fig04} MSD and CR-MSD of a hard disk system for various system sizes at $\phi = 0.81$ for $N = 1000$ up to $16000$ particles. While the CR-MSD (blue) does not show finite size effects in the plateau, the standard MSD (red) is strongly affected. Increasing opacity of the curves corresponds to increasing size. $250$ independent simulations were performed for each curve (except for $N=16000$ with $191$ runs). The inset plot shows the linear amplitude of the fluctuations $\sqrt{\mathrm{MSD}(\tau_i)}$ versus system size $L \propto \sqrt{N}$ in log-lin scale. $\tau_i$ is given by the inflection point of the curves, marked as gray circles. The straight line is a linear fit of the $\mathrm{log}_{10}(L)$ behaviour and verifies Mermin-Wagner fluctuations in 2D amorphous solids.
}
\end{figure}

In Figure~\ref{Fig03}, a separation of timescales between standard structural and orientational decay was shown for the 2D glasses. However, the $\alpha$-relaxation is strongly dependent on the supercooling but only marginally affected by system size. Escaping the cage is a local mechanism. Note that all experimental systems are much larger than the examined fields of view. The amplitude of Mermin-Wagner fluctuations on the other hand depends on elasticity (which is a function of temperature) but more importantly it depends logarithmicaly on system size \cite{Mermin1966,Mermin1968,Frohlich1981}. No predictions exist for the time scale of Mermin-Wagner fluctuations but it is reasonable to assume that they also depend on system size. Accidentally it might be the case that Mermin-Wagner fluctuations and $\alpha$-relaxation fall on top of each other. Therefore we vary systematically the number of particles for the simulated hard disk system at fixed packing fraction between $1000$ and $16000$ disks. Figure~\ref{Fig04} shows a comparison of cage-relative and normal MSD. All displacements measured in local coordinates collapse. The normal mean squared displacement on the other hand shows a strong finite size effect. The height of the plateau is extracted by taking the amplitudes of the displacement at the inflection point $\langle r^2(\tau_i)\rangle$ indicated by open circles in Fig.~\ref{Fig04}, being determined by fitting a 3rd-order-polynomial in the region of interest. Note, that the inflection point shifts to later times for larger systems, validating the assumption above.

Plotting the square root of the amplitude of the inflection point as function of the logarithm of the linear system size $L\propto \sqrt{N}$ gives a straight line. This is the logarithmic fingerprint of Mermin-Wagner fluctuations. A recent manuscript by H. Shiba et. al \cite{Shiba2016} independently reports the same logarithmic increase of long wavelength fluctuations in 2D of a simulated soft sphere glass, while to our knowledge they have not yet been directly measured in crystals.

\section{Discussion}

In \cite{Flenner2015} E. Flenner and G. Szamel reported fundamental differences between glassy dynamics in two and three dimensions and detected strong finite size effects in 2D. The localization in a 2D glass, e.g. measured by the plateau of the intermediate scattering function or mean squared displacement is not well pronounced and decays faster in large systems. Bond-order correlation functions (taking only six-folded particles into account) decay later and show less size dependence. Additionally particle trajectories show sudden jumps in 3D but not in 2D and dynamical heterogeneities are significantly pronounced in 3D. They conclude, that vitrification in 2D and 3D is not the same calling for a re-examination of the present glass transition paradigm in 2D. This re-examination is done by taking Mermin-Wagner fluctuations into account, the observed differences disappear using local coordinates \cite{Keim2015}. E. Flenner and G. Szamel investgated remarkably large systems to reduce finite size effects but this even enhances Mermin-Wagner fluctuations: those fluctuations affect translational degrees of freedom but not orientational ones and depend logarithmicaly on the system size. In \cite{Mazoyer2009} we showed that non-Gaussian behavior of the self part of the van-Hove function (which measures the variance of displacements at a given time) in 2D systems is only visible in local, cage-relative coordinates. Equivalently the dynamical heterogeneities show significantly more contrast in cage-relative coordinates. Mermin-Wagner fluctuations simply 'smear out' local events like hopping and cage-escape if particles are measured in a global coordinate frame \cite{Mazoyer2009}.

In \cite{Mazoyer2011} we assumed that the presence of collective motion might be due to long wavelength fluctuations. Now this is validated in direct comparison with 2D crystals. A recent manuscript by S. Vivek et al. \cite{Vivek2016} reports results of a soft sphere and a hard sphere glass in 2D compared to a 3D glass. Using similar correlation functions, their results are essentially the same but the (almost) hard sphere glass showed less signature of Mermin-Wagner fluctuations. This can be explained by the work of Fr\"ohlich et al. \cite{Frohlich1981}: they determined the following conditions for Mermin-Wagner fluctuations to appear: i) the pair-potential of particles has to be integrable in the far field and ii) analytically at the origin. The first condition rules out Coulomb interaction, since for this long range potential the second, third, and higher nearest neighbours interaction is strong enough that particle displacements can not add up statistically independent. The second condition questions hard sphere interaction. An easy argument in the limit of zero temperature might go as follows; when all particles are at contact and closed packed, no positional fluctuations can appear at all. At finite temperature the Mermin-Wagner fluctuations are excited as shown in Fig.~\ref{Fig04} but the separation of timescales is less pronounced in Fig.~\ref{Fig03} for the hard discs simulation, consistent with the results of Vivek et. al \cite{Vivek2016}. An alternative ansatz to investigate Mermin-Wagner fluctuations is reported by H. Shiba and coworkers, showing analogue results in large scale computer simulations. Shiba et al. analysed bond-breakage correlations, four-point correlations \cite{Shiba2012}, and intermediate scattering functions in 2D and in 3D. Being a local quantity, bond-breakage correlations do not differ in 2D and 3D thus the microscopic nature of the glass transition is similar. The density of vibrational states on the other hand of a 2D system computed from the velocity autocorrelation function shows an infinite growth of acoustic vibrations, very similar to 2D crystals \cite{Shiba2016}. Those beautiful results are completely in line with our arguments.

Connecting Mermin-Wagner fluctuations and glassy behaviour in 2D points to another question which is yet not completely solved, namely how (shear) solidity appears. Glass is a solid on intermediate time scales but on infinite times it may flow. A 2D crystal is soft on infinite length scales due to the lack of global spontaneous symmetry breaking and a solid only at intermediate scales.

\section{Conclusions}

Comparing experimental and simulation data in 2D and 3D, we show that a 2D crystal is affected by Mermin-Wagner-fluctuations in the long time limit, a 2D glass has Mermin-Wagner fluctuations as a 'second channel of decay' beside the standard $\alpha$-process of structural relaxation, while a 3D glass only shows an $\alpha$-process.  The existence of Mermin-Wagner fluctuations is not limited to low dimensional \emph{crystals}; they also appear in \emph{2D amorphous solids} and certainly in 2D quasi-crystals. The comparison of structural and orientational measures shows that Mermin-Wagner fluctuations can explain the differences in orientational relaxation times (not affected by long wavelength fluctuations) and translational relaxations times (affected by those fluctuations). Furthermore, Mermin-Wagner fluctuations exist on large scales and the local effect (within the size of the cage) is only an affine translation. Thus the cage-escape mechanisms is not influenced by Mermin-Wagner fluctuations. We can conclude that the microscopic mechanism of the 2D and 3D glass transition is not necessarily different while the transient localization measured by global variables is less pronounced in 2D compared to 3D.\\

P.K. acknowledges fruitful discussion with Herbert Wagner and financial support from the Young Scholar Fund, University of Konstanz. H.K. acknowledges the synthesis of PMMA colloids by M.K. Klein and A. Zumbusch.

\bibliographystyle{apsrev4-1}
\bibliography{Kolloide}

\end{document}


\title{Supplemental information: Mermin-Wagner fluctuations in 2D amorphous solids}

\author{Bernd Illing}
\affiliation{Department of Physics, University of Konstanz, D-78457 Konstanz, Germany}
\author{Sebastian Fritschi}
\affiliation{Department of Physics, University of Konstanz, D-78457 Konstanz, Germany}
\author{Herbert Kaiser}
\affiliation{Department of Physics, University of Konstanz, D-78457 Konstanz, Germany}
\author{Christian Klix}
\affiliation{Department of Physics, University of Konstanz, D-78457 Konstanz, Germany}
\author{Georg Maret}
\affiliation{Department of Physics, University of Konstanz, D-78457 Konstanz, Germany}
\author{Peter Keim}
\email{peter.keim@uni-konstanz.de}
\affiliation{Department of Physics, University of Konstanz, D-78457 Konstanz, Germany}

\date{\today}

\begin{abstract}
Here we give additional information about the experiments and the simulation and briefly discuss how pinning of particles at solid substrates influences Mermin-Wagner fluctuations.
\end{abstract}


\maketitle

The confinement of colloids to two dimensions is achieved by sedimentation to an interface which in our case is the lower water/air surface in hanging dropled geometry. The volume is regulated actively with a microsyringe to get a flat interface. The fluid interface guarantees free diffusion of the colloids (without any pinning) within the horizontal monolayer but the sedimentation height is about $20~\mathrm{nm}$ and negligible compared to the size of the colloids. The latter consist of polystyrene beads with diameter $\sigma_A=4.5\;\mu\textrm{m}$ (species A) for the crystals and a mixture of species A and B (to avoid crystallization) for the glass samples. Species B has $\sigma_B = 2.8\,\mu m$ and the glassy mixture has a relative concentration of $\xi = N_B/(N_A+N_B)\approx 50\,\%$ where $N_A$ and $N_B$ are the number of particles of both species in the field of view. The colloidal beads are further doped with iron oxide nano-particles which results in a super-paramagnetic behavior and a mass density of 1.7 kg/dm$^3$. The whole monolayer consists of several hundred thousand particles where about $\approx2000$ particles are monitored by video-microscopy in a $1158\times865\;\mu\textrm{m}^2$ sub window in the glass sample and $\approx3000$ particles in a $835\times620\;\mu\textrm{m}^2$ sub window in the crystalline sample. The individual colloids are tracked with a spatial resolution of $\approx 50~\mathrm{nm}$ and with a time resolution of the order of a second. The system is kept at room temperature and exempt from density gradients due to a month-long precise control of curvature and inclination of the interface.

\begin{figure}
\centering
\includegraphics[width=.5\textwidth]{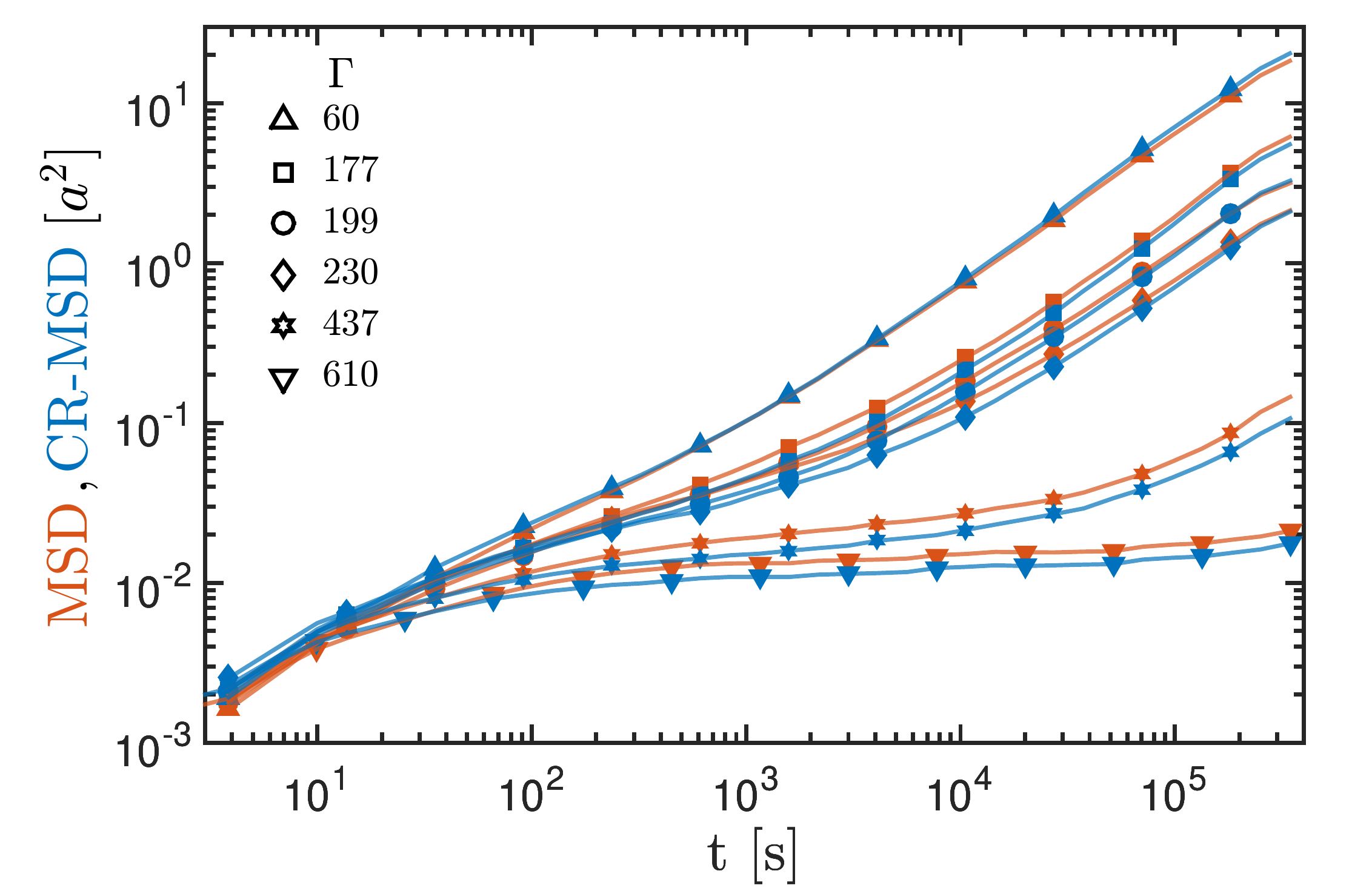}
\includegraphics[width=.5\textwidth]{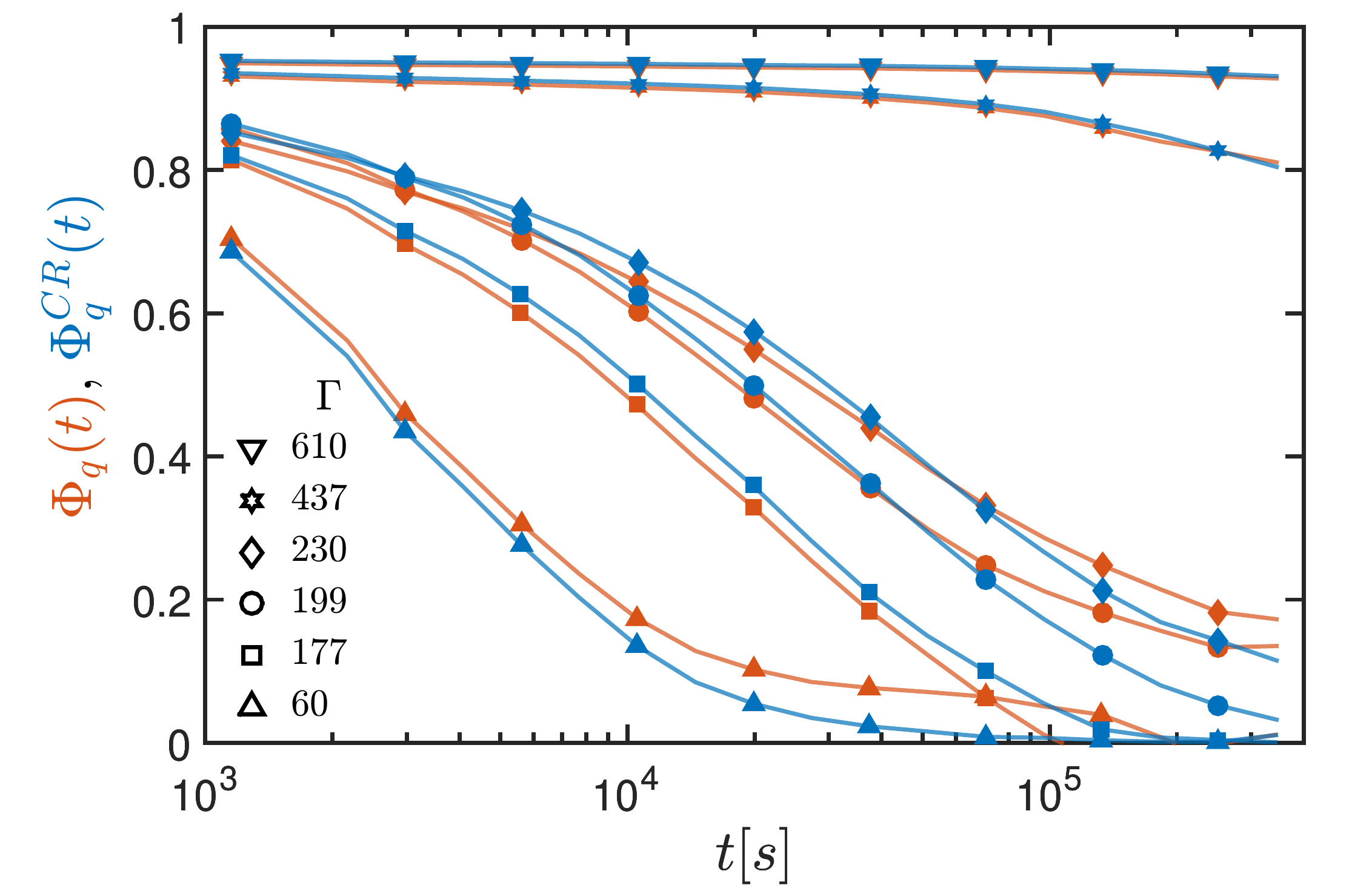}
\caption{Upper plot: Mean squared displacement of a colloidal monolayer on a solid substrate. Accidentally  between 1\% to 3\% of particles were pinned to the substrate which efficiently suppress Mermin-Wagner fluctuations. The upper two curves are fluid, the circles indicate a system close to the glass transition while the lower three curves are solid. The lower plot shows the density correlator (see main text) for the same system (again $qa=2\pi$).}
\label{SI_Fig01}
\end{figure}

Due to the super-paramagnetic nature of the particles, the potential energy can be tuned by means of an external magnetic field $\vec{H}$ applied perpendicular to the monolayer which induces a repulsive dipole-dipole interaction between the particles. The ratio between potential energy $E_{\textrm{mag}}$ and diffusive thermal energy $k_BT$,
\begin{equation}
    \Gamma = \frac{\mu_0}{4\pi} \cdot \frac{H^2 \cdot (\pi n)^{3/2}}{k_BT}(\xi \cdot \chi_B + (1-\xi)\cdot \chi_A)^2 \, ,
\label{eqnSI01}
\end{equation}
acts as inverse temperature (or dimensional pressure for fixed volume and particle number). $n=1/a^2$ is the 2D particle density with a mean particle distance $13~\mu\textrm{m} < a < 22~\mu\textrm{m}$ (depending on the kind of sample). The magnetic susceptibility per bead is $\chi_A=6.5\cdot10^{-11}\;\textrm{Am}^2/\textrm{T}$ and $\chi_B=6\cdot10^{-12}\;\textrm{Am}^2/\textrm{T}$ for species A and B respectively. For low $\Gamma$ the system is fluid since the thermal motion dominates. Increasing $\Gamma$ induces crystallization at $\Gamma_m = 70$ \cite{Deutschlaender2014} or dynamic arrest and vitrification at  $\Gamma_G = 195$ \cite{Klix2015}. Digital image analysis gives the positions of the beads as function of time. This provides the complete phase-space information of the colloidal ensembles at all relevant time scales on an 'atomic' level. Our glasses are 'well aged'; we carefully checked that the $\alpha$-process is independent of waiting time. Additional details of the 2D setup are described elsewhere \cite{Ebert2009a}.
Using the water/air interface is crucial: we performed additionally a whole set of measurements as function of $\Gamma$ on a solid substrate (object plate) where accidently between 1\% to 3\% of particles were pinned. This pinning sufficiently suppresses Mermin-Wagner fluctuations and the differences between MSD and CR-MSD disappears, as shown in Fig.~\ref{SI_Fig01}.\\

The 3D glass consists of monodisperse, charged polymethylmetacrylate (PMMA) particles of $\sigma_C = 2.6\;\mu\textrm{m}$ diameter, covered with poly-12-hydroxystearic acid (PHSA) to prevent aggregation which are dissolved in a well adjusted cyclohexylbromide (CHB) and decaline mixture. This mixture matches the mass density to prevent sedimentation and simultaneously the refractive index of solvent and colloid. Due to the latter the ensemble is not turbid and one can look deep into the sample (which consist of several billion particles). The particles are doped with rhodamine 6G and by using standard confocal microscopy, the 3D trajectories of up to $6000$ particles can be monitored and analysed simultaneously in a field of view of about $70 \times 60 \times 30\;\mu\textrm{m}^3$. Care has to be taken to use thoroughly cleaned solvents. CHB was passed through an activated alumina column to remove H$_2$O, HBr, and any other polar molecules and ions \cite{Pangborn1996}. Dispersed in CHB, PMMA colloids acquire a moderate positive charge. This seems to originate from the adsorption of protons and bromide ions stemming from the decomposition of CHB (with protons being adsorbed more likely) \cite{vanDerLinden2015}. Together with the small abundance of ions in the solvent one obtains a screened Coulomb interaction with the dimensionless Yukawa potential:
\begin{equation}
    u(r) = l_B \left(\frac{e^{\kappa/2}}{1+\kappa\sigma_C/2}\right)^2 Z^2_{\mathrm{eff}}\cdot \frac{e^{-\kappa r}}{r} \, ,
\label{eqnSI02}
\end{equation}

\vspace{-2mm}
\begin{table}[h]
\centering
\caption{\label{Tab01} Parameters of the 3D glass}
\begin{tabular}{|c|c|c|c|}
  \hline
            & $\Theta~[\%] $ & $\kappa^{-1}~[a_0]$ & $Z_{\mathrm{eff}}$ \\ \hline
  glass \#1 &    21.5         &      0.183           &     500 \\
  glass \#2 &    19.2         &      0.176           &     500 \\
  glass \#3 &    19           &      0.166           &     500 \\
  \hline
\end{tabular}
\end{table}
where $\kappa $ is the inverse screening length of the ions and $l_B = e^2 / (k_BT \cdot 4\pi \epsilon_0 \epsilon_r)$ is the Bjerrum length. This is the length, where the interaction potential between two elementary charges $e$ in the medium with dielectric constant $\epsilon_r$ equals the thermal energy $k_BT$ ($\epsilon_0$ is the dielectric constant of the vacuum). System parameters are determined from fits of theoretically computed structure factors $S(q)$ (using an approximative integral equation theory \cite{Heinen2011}) to the experimental data. The effective mean charge $Z_{\mathrm{eff}} = 500$ on the surface of the colloid is the same for all three presented systems. Thus, the phase diagram has two relevant parameters left which counteract; these are the packing fraction $\Theta$ which is given by the particle distance (and size) and the screening length $\kappa^{-1}$ given by the density of counterions and any other ions. The charge poly-dispersity $\Delta Z$ is dominated by the surface area poly-dispersity thus being of the order of up to $17\%$ for a size poly-dispersity determined by electron microscopy to be about $8\%$. This effectively suppresses crystallization which occurs not earlier than one to two weeks after rejuvenation of the sample by shaking. Table~\ref{Tab01} shows the parameters of three different glasses.\\

\begin{figure}
\centering
\includegraphics[width=.5\textwidth]{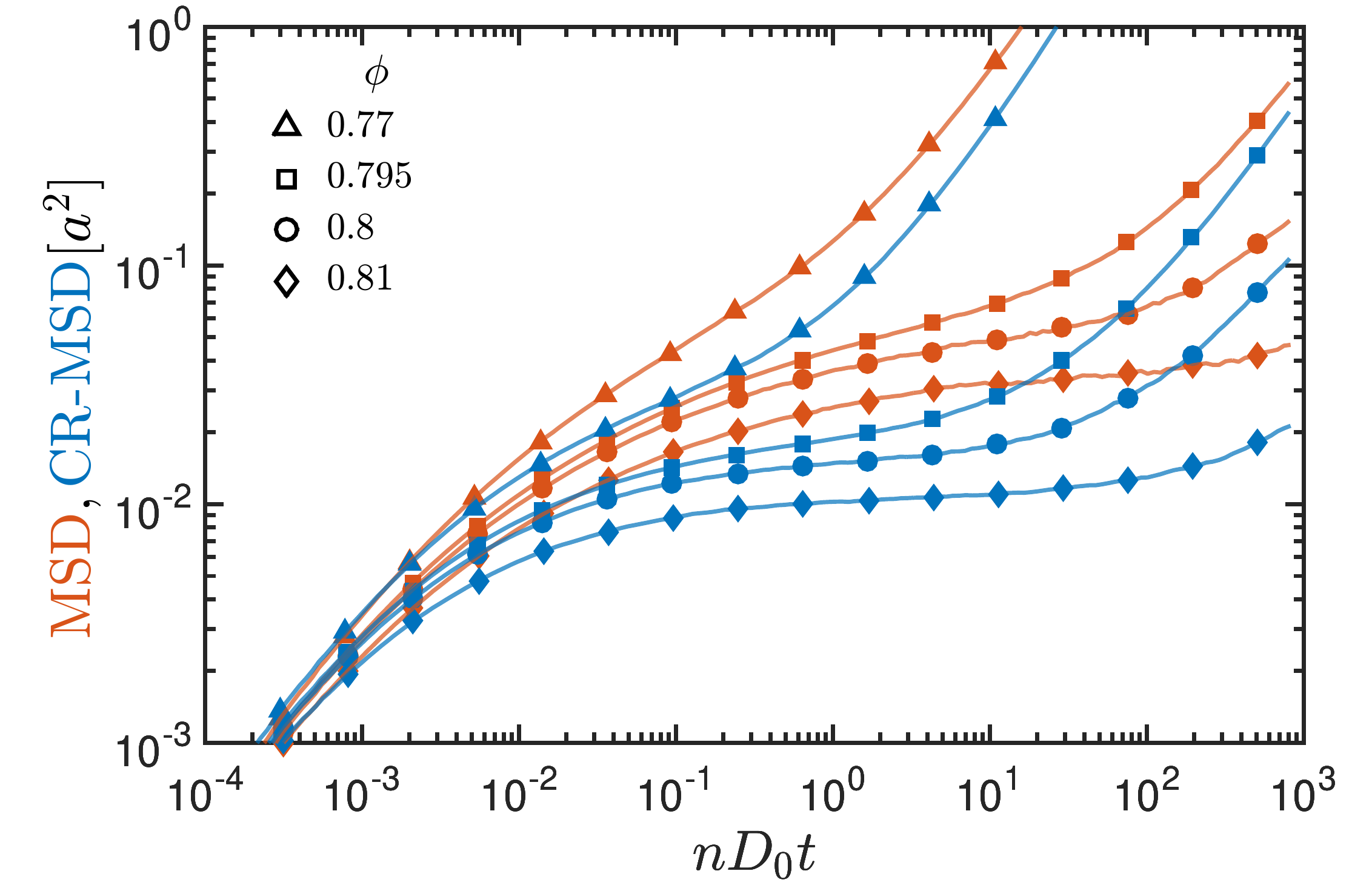}
\caption{The separation of MSD (red) and CR-MSD (blue) for the hard disc system shows Mermin-Wagner fluctuations. The triangles indicate a fluid system, the squares indicate a system with critical packing fraction while the circles and diamonds show a solid sample. The data are the same as for the density correlator in Fig.~[3] of the main manuscript except of the fluid data.}
\label{SI_Fig02}
\end{figure}

In the following we will also compare experimental results with simulations. Those are made for a 2D binary mixture of hard disks undergoing Brownian motion employing an event-driven simulation algorithm \cite{Scala2007,Henrich2009}. The system contains $N=16000$ particles, is made up of a 50:50 mixture with diameters $d_{\mathrm{A}} = 1$, $d_{\mathrm{B}} = 1.4$, and is equilibrated by Newtonian dynamics before data are collected. The packing fraction $\phi$, giving the ratio of the area occupied by the disks to the area of the system, varies from $\phi=0.77$ to $\phi=0.81$, hence in the vicinity of the glass transition point $\phi_{\mathrm{c}} \approx 0.795$ \cite{Weysser2011}. For the simulations the averaging (angular brackets) is done for typically $100$ runs as function of density and for finite size effects up to $200$ independent runs as function of system size. Fig.~\ref{SI_Fig02} shows the comparison of MSD and CR-MSD in similar manner as Fig.~[2] of the main manuscript for the experimental systems.

We observed Mermin-Wagner fluctuations in 2D crystal and 2D glass. This rises the question which low dimensional systems will show Mermin-Wagner fluctuations? Since well defined neighbour distances with low variance as in glasses, quasi-crystals, and crystals are required, we suggest hyperuniformity, originally formulated to characterize structures with isotropic photonic bandgaps \cite{Florescu2009,Man2013}, to be a necessary requirement: in hyperuniform structures the number variance $\sigma(R)= \langle N_R^2\rangle - \langle N_R \rangle^2$ of $N_R$ atoms within a n-dimensional sphere of radius $R$ is proportional to the n-1 dimensional surface of the sphere.

\bibliographystyle{apsrev4-1}
\bibliography{Kolloide}